\documentstyle[11pt]{article}
\input boxedeps.tex
\input epsf
\input colordvi
\SetRokickiEPSFSpecial
\HideDisplacementBoxes
\begin{document}
\pagestyle{empty}
\begin{flushright}
\tt LATECH-CAPS-99-09a
\end{flushright}
\vspace{0.5cm}
\begin{flushleft}
\ForceWidth{\textwidth}\BoxedEPSF{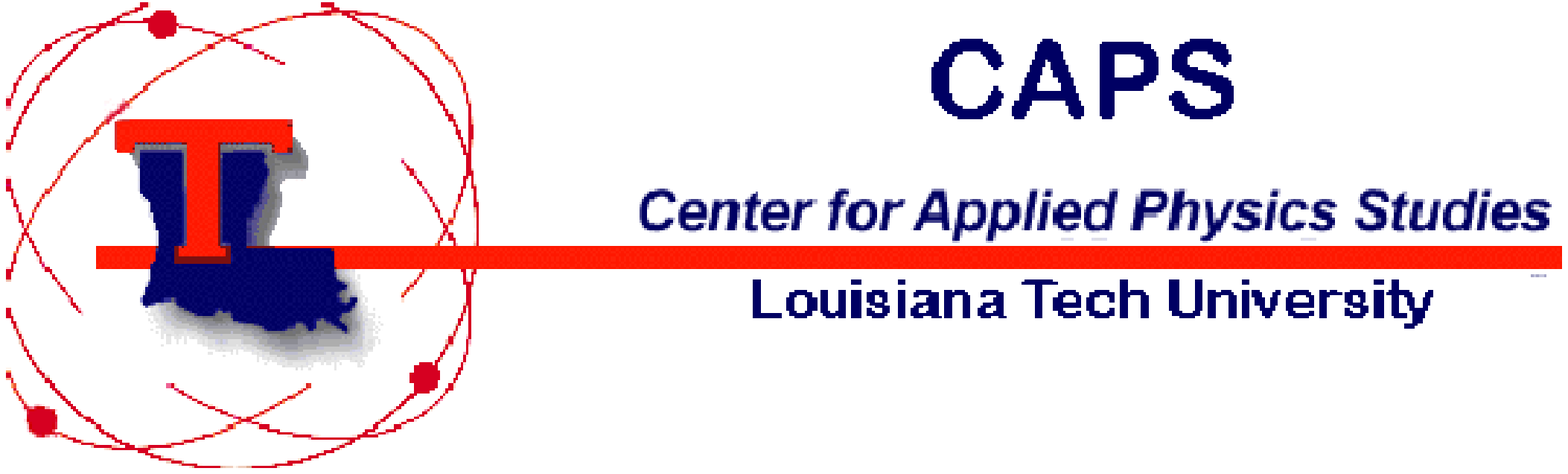}
\end{flushleft}
\vspace{2.5cm}

\begin{center}
{\LARGE Implications of Weak-Interaction Space Deformation for Neutrino Mass Measurements} 

\medskip
\medskip
\medskip

{\Large Neven Simicevic}                            

{\large\it Louisiana Tech University}\\           
{\large\it Center for Applied Physics Studies}\\  
{\large\it Ruston, LA 71272}\\                          
{\large\today}\\                                  

\bigskip

\end{center}

\newpage
\pagestyle{plain}

\begin{abstract}

The negative values for the squares of both electron and muon neutrino masses 
obtained in recent experiments are explained as a possible consequence of a 
change in metric within the weak-interaction volume in the energy-momentum 
representation. Using a model inspired by a combination of the general theory of 
relativity and the theory of deformation for continuous media,
it is shown that the negative value of the square of the neutrino mass can be 
obtained without violating allowed physical limits. The consequence is that the 
negative value is not necessary unphysical. 
 
\end{abstract}

\section{Introduction}

There has recently been some concern over the significant negative 
values obtained for the square of the neutrino mass, for both the electron neutrino 
\footnote{We will refer to both neutrino and antineutrino just as neutrino.}
measured in nuclear $\beta$-decay, and the muon neutrino 
measured  in ${\pi^{+}} \to {\mu^{+}}{\nu_{\mu}}$ decay \cite{Part94}.
In these measurements, the probability that the square of the electron neutrino 
mass $m_{\nu_{e}}^2$ is positive is only $3\%$ \cite{Part94}, while from most 
recent measurements the square of the muon neutrino mass $m_{\nu_{\mu}}^2$ 
is negative by 6.1 or 0.9 standard deviations, depending on the choice of 
the pion rest mass value \cite{Part94,Assamagan94}. 

While it may be argued that the negative values obtained for  $m_{\nu_{e}}^2$
and $m_{\nu_{\mu}}^2$ are a consequence of systematic errors in these 
measurements which are still not understood, we must also investigate the 
possibility that there is a physical underlay for the measured results. 
In this paper, we therefore propose a mechanism which, in principle, allows 
the measured square of the neutrino mass to be negative, while at the same 
time does not cross allowed physical mass limits. Our assumption is that the 
negative values of $m_{\nu_{e}}^2$ and $m_{\nu_{\mu}}^2$ are not 
consequences of the dynamics of the decay, but rather are the result of the 
geometry, or  metric, of the small volume in which the weak interaction drives 
the decay.

The large masses of the intermediate vector bosons $W^{+}$, $W^{-}$ and $Z^{0}$
result in a very short range for the weak interaction. The interaction
volume is small enough to ensure the success of the Fermi $\beta$-decay theory
\cite{Fermi37},
in which the interaction is assumed to be a four-particle coupling. One may
expect that the metric valid in vacuum is not necessarily 
valid in a volume with such a small dimension, especially
considering that we already have a change in vacuum metric in 
the presence of mass, in accordance with
the general theory of relativity.

In this paper we study the consequence of the change of metric
within the weak-interaction volume on the measured value of the square 
of the neutrino mass. In our model, we make minimal changes to the metric, 
in a manner which is as simple as possible, and only as much as necessary 
to make it different from the vacuum metric, yet at the same time have 
clear physical consequences.

\section{General basis for the model}

The basic idea of this model is essentially the same as that of the general 
theory of relativity. Translated for the purpose of this paper, it means that 
the metric of space deforms at the distance scale comparable to the range of 
the weak interaction. The majority of the  necessary mathematical formalism 
can be taken from the theory of the deformation of continuous media, and 
extended to 4-dimensional Minkowski space. We  can explore the consequences 
of this space deformation without having the necessity to propose 
the exact mechanism which causes the deformation.
      
We define two 4-dimensional Minkowski spaces, an undeformed space $C$ and 
a deformed space $C^{*}$. Let us choose a point 
$P({\xi^{0}},{\xi^{1}},{\xi^{2}},{\xi^{3}})$ in the undeformed space $C$. 
Under the deformation of the space $C$, the point 
$P({\xi^{0}},{\xi^{1}},{\xi^{2}},{\xi^{3}})$ from the undeformed space is mapped 
into the point $P^{*}({\xi^{*0}},{\xi^{*1}},{\xi^{*2}},{\xi^{*3}})$ of the deformed 
space $C^{*}$. The space deformation is defined by the equations analogous to 
the Lagrangian coordinate point of view \cite{Sokolnikoff58,Boresi65}:

\begin{eqnarray}
\xi^{*0}=\xi^{*0}({\xi^{0}},{\xi^{1}},{\xi^{2}},{\xi^{3}}); \; \;  
\xi^{*1}=\xi^{*1}({\xi^{0}},{\xi^{1}},{\xi^{2}},{\xi^{3}}); \nonumber \\
\xi^{*2}=\xi^{*2}({\xi^{0}},{\xi^{1}},{\xi^{2}},{\xi^{3}}); \; \;  
\xi^{*3}=\xi^{*3}({\xi^{0}},{\xi^{1}},{\xi^{2}},{\xi^{3}}),   \label{deff}
\end{eqnarray}
or the equations analogous to the Euler coordinate point of view:
\begin{eqnarray}
\xi^{0}=\xi^{0}({\xi^{*0}},{\xi^{*1}},{\xi^{*2}},{\xi^{*3}}); \; \;  
\xi^{1}=\xi^{1}({\xi^{*0}},{\xi^{*1}},{\xi^{*2}},{\xi^{*3}}); \nonumber \\
\xi^{2}=\xi^{2}({\xi^{*0}},{\xi^{*1}},{\xi^{*2}},{\xi^{*3}});  \; \; 
\xi^{3}=\xi^{3}({\xi^{*0}},{\xi^{*1}},{\xi^{*2}},{\xi^{*3}}).  \label{deff2}
\end{eqnarray}
These functions must be continuous and differentiable in the space $C$ ($C^{*}$), 
because a discontinuity in these functions would imply a ``rupture'' of the space 
$C$ ($C^{*}$). 

Following the classical theory of elasticity \cite{Sokolnikoff58,Boresi65} we 
define the infinitesimal length $ds$ as a line element $PQ$ between two points
$P({\xi^{0}},{\xi^{1}},{\xi^{2}},{\xi^{3}})$ and 
$Q({\xi^{0}}+d{\xi^{0}},{\xi^{1}}+d{\xi^{1}},{\xi^{2}}+d{\xi^{2}},{\xi^{3}}+d{\xi^{3}})$,
and the infinitesimal length $ds^{*}$ as a line element $P^{*}Q^{*}$ between 
two points $P^{*}({\xi^{*0}},{\xi^{*1}},{\xi^{*2}},{\xi^{*3}})$ and 
$Q^{*}({\xi^{*0}}+d{\xi^{*0}},{\xi^{*1}}+d{\xi^{*1}},{\xi^{*2}}+d{\xi^{*2}},
{\xi^{*3}}+d{\xi^{*3}})$.
Under the deformation, the length $ds$ can either be elongated or contracted. 
The magnification of the deformation is defined as $ds^{*} \over ds$.
In the undeformed space $C$, $ds$ is defined as the square root of the invariant 
form

\begin{equation}
{ds^{2}} = {g_{\mu\nu}}{d{\xi^{\mu}}}{d{\xi^{\nu}}},   \label{metten}
\end{equation}
where $g_{\mu\nu}$ are elements of a symmetric tensor $[g_{\mu\nu}]$ defining 
the metric of the space  $C$.
The metric tensor $[g_{\mu\nu}]$,

\begin{equation}
{[g_{\mu\nu}]}=\left [\begin {array}{cccc} 1&0&0&0\\0&-1&0&0\\0&0&-1&0\\
0&0&0&-1
\end {array}\right ], \label{metmin}
\end{equation}
defines four-dimensional
Minkowski space in the orthogonal cartesian representation in the system 
where the speed of light $c=1$. The distance between two points 
$P({t},{x},{y},{z})$ and $Q({t+dt},{x+dx},{y+dy},{z+dz})$,  where $t$ represents 
time and $(x,y,z)$ space, is the square root of the invariant form 
\begin{equation}
{ds^{2}}={dt^{2}}-{({dx^{2}}+{dy^{2}}+{dz^{2}})}. \label{inform}
\end{equation}
In the deformed space $C^{*}$, $ds^{*}$ is the square root of the invariant form
\begin{equation}
{ds^{*2}} = {g_{\mu\nu}}{d{\xi^{*\mu}}}{d{\xi^{*\nu}}}.   \label{mett}
\end{equation}
In the deformed Minkowski  space defined
by time  $t^{*}$ and space coordinates $(x^{*},y^{*},z^{*})$, the distance 
between two points $P^{*}({t^{*}},{x^{*}},{y^{*}},{z^{*}})$ and 
$Q^{*}({t^{*}+dt^{*}},{x^{*}+dx^{*}},{y^{*}+dy^{*}},{z^{*}+dz^{*}})$
is the square root of the invariant form 
\begin{equation}
{ds^{*2}}={dt^{*2}}-{({dx^{*2}}+{dy^{*2}}+{dz^{*2}})}. \label{inform2}
\end{equation}

Using again general notation, we can express total differentials 
$(d{\xi^{*0}},d{\xi^{*1}},d{\xi^{*2}},d{\xi^{*3}})$ in terms of total differentials 
$(d{\xi^{0}},d{\xi^{1}},d{\xi^{2}},d{\xi^{3}})$ by the transformation
\begin{equation}
d\xi^{*\nu}= a^{\nu\mu}d\xi^{\mu},   \label{tr}
\end{equation}
where 
\begin{equation}
a^{\nu\mu}={\partial\xi^{*\nu} \over \partial\xi^{\mu}}. \label{tra}
\end{equation}
The measure of deformation  $ds^{*2}-ds^{2}$ can be then calculated as

\begin{eqnarray}
ds^{*2}-ds^{2} &=& g_{\mu\nu}(d\xi^{*\mu}d\xi^{*\nu}-d\xi^{\mu}d\xi^{\nu}) 
\nonumber \\
 &=& (g_{\iota\kappa}a^{\iota\mu}a^{\kappa\nu}-g_{\mu\nu})d\xi^{\mu}d\xi^{\nu} 
 \nonumber \\
 &=& \epsilon_{\mu\nu}d\xi^{\mu}d\xi^{\nu} \label{dif1}
\end{eqnarray}
or
\begin{eqnarray}
ds^{*2}-ds^{2} &=& g_{\mu\nu}(d\xi^{*\mu}d\xi^{*\nu}-d\xi^{\mu}d\xi^{\nu}) 
\nonumber \\
 &=& (g_{\mu\nu}-g_{\iota\kappa}b^{\iota\mu}b^{\kappa\nu})d\xi^{*\mu}d\xi^{*\nu} 
 \nonumber \\
 &=& \eta_{\mu\nu}d\xi^{*\mu}d\xi^{*\nu}\label{dif2}
\end{eqnarray}
where 
\begin{equation}
b^{\nu\mu}={\partial\xi^{\nu} \over \partial\xi^{*\mu}}. \label{tratra}
\end{equation}

To simplify the calculation, we can rotate our coordinate system such that 
the coordinate axes correspond to the principal directions of the deformation 
tensors $\epsilon=[\epsilon_{\mu\nu}]$ and  $\eta=[\eta_{\mu\nu}]$. The new 
coordinate system corresponds to orthogonal directions in the undeformed 
space which remain orthogonal after deformation \cite{Sokolnikoff58,Boresi65}. 
In this case, the quadratic forms in Eq.'s ~\ref{dif1}~  and ~\ref{dif2}~ reduce to 
their canonical forms, and the deformation tensors have diagonal form: 
$\epsilon=[\epsilon_{\mu\mu}]$ and  $\eta=[\eta_{\mu\mu}]$.
The deformation $ds^{*2}-ds^{2}$ is now

\begin{equation}
ds^{*2}-ds^{2}=\epsilon_{\mu\mu}d\xi^{\mu}d\xi^{\mu} \label{dif3}
\end{equation}
or
\begin{equation}
ds^{*2}-ds^{2}=\eta_{\mu\mu}d\xi^{*\mu}d\xi^{*\mu}.\label{dif4}
\end{equation}

We are now in the same position as in the general theory of relativity 
where the existence of the gravitational potential changes 
the metric tensor \cite{Sokolnikoff58,Bergmann42,Anderson67,Adler75},  
whose coefficients become functions
of the local coordinates and can be written in the general form 
\begin{equation}
{[g_{\mu\nu}^{'}]}=\left [\begin {array}{cccc} {f_{00}}({\xi})&0&0&0
\\0&{f_{11}}({\xi})&0&0
\\0&0&{f_{22}}({\xi})&0
\\0&0&0&{f_{33}}({\xi})
\end {array}\right ]. \label{gp}
\end{equation}
$f_{\mu\nu}({\xi})=f_{\mu\nu}({\xi^{0}},{\xi^{1}},{\xi^{2}},{\xi^{3}})$ 
are functions generating the deformation of four-dimensional space. 

In this paper we address the effect of this space deformation on the
kinematics in the deformed region. Generally, we can again define two spaces, 
an undeformed space $\cal C$ and a deformed space ${\cal C}^{*}$. Under 
the deformation, the point ${\cal P}({\pi^{0}},{\pi^{1}},{\pi^{2}},{\pi^{3}})$ in the 
undeformed space $\cal C$, is mapped into the point 
${\cal P}^{*}({\pi^{*0}},{\pi^{*1}},{\pi^{*2}},{\pi^{*3}})$ of the deformed 
space ${\cal C}^{*}$. The invariant form is here defined as

\begin{equation}
{dm^{2}} = {g_{\mu\nu}}{d{\pi^{\mu}}}{d{\pi^{\nu}}}.   \label{metten2}
\end{equation}
Translated into  the orthogonal cartesian 
energy-momentum representation $(E,{p_{x}},{p_{y}},{p_{z}})$,
the distance between two points ${\cal P}(E,{p_{x}},{p_{y}},{p_{z}})$ and
${\cal Q}(E+dE,{p_{x}+dp_{x}},{p_{y}+dp_{y}},{p_{z}+dp_{z}})$  is the 
square root of the same form as in Eq.~\ref{inform}, i.e.,
\begin{equation}
{dm^{2}}={dE^{2}}-{({dp_{x}^{2}}+{dp_{y}^{2}}+{dp_{z}^{2}})}. \label{infep}
\end{equation}
The effect of the space deformation is analogous to the effect 
in Eq.'s ~\ref{dif3}~  and ~\ref{dif4}~,

\begin{equation}
dm^{*2}-dm^{2}={\cal E}_{\mu\mu}d\pi^{\mu}d\pi^{\mu} \label{difp3}
\end{equation}
and
\begin{equation}
dm^{*2}-dm^{2}={\cal G}_{\mu\mu}d\pi^{*\mu}d\pi^{*\mu}. \label{difp4}
\end{equation}
The transformation from the space  $\cal C$ into the space ${\cal C}^{*}$ can 
be found in analogy with Eq.'s ~\ref{tr}~ and ~\ref{tra}~. To relate the 
transformation coefficients, we postulate that the Heisenberg relations hold 
even if the space is deformed, and that the number of possible states cannot 
be increased or decreased by the mechanism causing space deformation. 
This means that, for each index $\mu$,
\begin{equation}
d\xi^{\mu}d\pi^{\mu}=d\xi^{*\mu}d\pi^{*\mu}. \label{hx}
\end{equation}
As a result, the transformation is
\begin{equation}
d\xi^{*\mu}d\pi^{*\mu}= a^{\mu\nu}d\xi^{\nu}b^{\mu\nu}d\pi^{\nu}
=\delta^{\mu\nu}d\xi^{\nu}d\pi^{\nu}
= d\xi^{\mu}d\pi^{\mu}, \label{hx2}
\end{equation}
obviously  $b^{\mu\mu}=(a^{\mu\mu})^{-1}$, and:
\begin{equation}
d\pi^{*\mu}= b^{\mu\mu}d\pi^{\mu};  \; \;  d\pi^{\mu}= a^{\mu\mu}d\pi^{*\mu}. 
\label{trtr}
\end{equation}
The coefficients $a^{\mu\mu}$ and $b^{\mu\mu}$ are defined in Eq.'s ~\ref{tra}~  
and ~\ref{tratra}.

\section{Application to the neutrino mass measurement}
 
In this section we study the consequences of the general formalism developed in 
the previous section for the neutrino mass measurements.  Because the claim 
of this paper is that the negative values of the squares of the electron and muon 
neutrino masses are a consequence of the change in metric in the 
weak-interaction volume, the same model should be applied for both types of 
neutrino. There are many ways to deform the volume, but  for the sake 
of simplicity we choose a very simple  model of deformation, with just one 
free parameter.  The model is a copy of the mathematical formalism of a 
mechanical deformation caused by hydrostatic pressure into our 
geometry \cite{Boresi65}. We also assume that the ``time'' (or ``energy'') 
component is not affected.  In this case, our deformation tensor 
is given by

\begin{equation}
{[{\cal E}_{\mu\mu}]}=\left 
[\begin {array}{cccc} 1&0&0&0
\\0&-\epsilon&0&0\\0&0&-\epsilon&0\\0&0&0&-\epsilon
\end {array}\right ], \label{dten}
\end{equation}
where $\epsilon$ is a constant. We note here that in the limit 
$\epsilon \rightarrow 1$, $[{\cal E}_{\mu \nu}] \rightarrow [g_{\mu \nu}]$, 
and the undeformed Minkowski space is recovered. This choice of transformation
results in
 
\begin{equation}
{m_{\nu}^{*2}}={E^{*2}}-{({p^{*2}_{x}}+{p^{*2}_{y}}+{p^{*2}_{z}})}
={E^{2}}-{\epsilon ({p^{2}_{x}}+{p^{2}_{y}}+{p^{2}_{z}})}. \label{inene}
\end{equation}

The square of the neutrino masses measured in nuclear $\beta$-decay 
and in ${\pi^{+}} \to {\mu^{+}}{\nu_{\mu}}$ decay are
reconstructed using  energy and momentum conservation assuming a free space 
metric. If space is actually deformed, but not taken into account, the free 
undeformed space metric can result in negative values for the squares of the 
neutrino masses. Because two separate experimental programs  determine the 
electron neutrino and muon neutrino masses, while our model should describe 
both cases, we will use the results from the muon neutrino experiments to 
determine the parameter $\epsilon$, and then use that value
to see the consequences for the case of the electron neutrino measurements.

The muon neutrino mass can be measured from the decay reaction 
${\pi^{+}} \to {\mu^{+}}{\nu_{\mu}}$ \cite{Part94,Assamagan94}. 
For ${\pi^{+}}$ decay at rest in undeformed space, the muon neutrino mass 
is determined from the kinematic relation

\begin{equation}
m_{\pi}=\sqrt{m^{2}_{\nu_{\mu}}+p^{2}_{\mu}}+\sqrt{m^{2}_{\mu}+p^{2}_{\mu}},
\label{lenmfre}
\end{equation}
where momentum conservation has been imposed.
If the space is deformed, however, this relation becomes
\begin{equation}
m_{\pi}=\sqrt{m^{*2}_{\nu_{\mu}}+p^{*2}_{\mu}}
+\sqrt{m^{*2}_{\mu}+p^{*2}_{\mu}}.
\label{lenmfre2}
\end{equation}
Assuming that the ``true'' neutrino mass is zero
\footnote{This assumption, even if not critical for the discussion in this paper, 
does have a basis in the experiments which deal with ``free'' neutrinos,  
such as neutrino oscillation experiments, and suggest that the neutrino mass is 
very close to zero [9].}, then $m^{*2}_{\nu_{\mu}}=0$, and using the metric 
tensor defined in Eq.~\ref{dten}, we can calculate the value of the parameter 
$\epsilon$ from  Eq.'s ~\ref{lenmfre}~ and ~\ref{lenmfre2}~via

\begin{equation}
\sqrt{m^{2}_{\nu_{\mu}}+p^{2}_{\mu}}=p^{*}_{\mu}=\epsilon p_{\mu}.
\label{lenmfre3}
\end{equation}
One must notice that 
$\sqrt{m^{2}_{\mu}+p^{2}_{\mu}}=\sqrt{m^{*2}_{\mu}+p^{*2}_{\mu}}$ 
because of Eq.~\ref{dten}. The value of $\epsilon$ is thus determined by 
the momentum of muons from the pion decay, measured to be 
$p_{\mu}=29.79207 \pm 0.00024 $~$MeV/c$ \cite{Part94,Assamagan94}.

There are two solutions corresponding to two choices of the pion mass, which 
have been  labeled Solution A and Solution B \cite{Part94,Assamagan94}.
Solution A, for which $m_{\pi}=139.56782 \pm 0.00037$~$MeV$ 
and $m^{2}_{\nu_{\mu}}=-0.143 \pm 0.024$~$MeV^{2}$ \cite{Part94} 
yields for the parameter  $\epsilon$ a value
\begin{equation}
\epsilon=0.999988 \pm 0.000004, 
\label{epsa}
\end{equation}
while  Solution B with $m_{\pi}=139.56995 \pm 0.00035$~$MeV$ 
and $m^{2}_{\nu_{\mu}}=-0.016 \pm 0.023$~$MeV^{2}$ yields
\begin{equation}
\epsilon=0.9999998 \pm 0.0000004. 
\label{epsb}
\end{equation}
As should have been expected, the value of the parameter  $\epsilon$ is 
very close to 1, suggesting that the space  deformation is not very large.

We now apply the same model to the measurement of the electron neutrino 
rest mass. The electron neutrino rest mass determined from tritium 
$\beta$-decay is obtained from the shape of 
the $\beta$-spectrum close to the end-point, expressed as

\begin{equation}
W(E)=A F p (E+m_{e}) \sum_{i} W_{i} (E_{0i}-E) 
\sqrt{ (E_{0i}-E)^{2} - m_{\nu_{e}}^{2}},
\label{neuph}
\end{equation}
where $A$ is an amplitude, $F$ is the Fermi function, $m_{e}$ 
and $m_{\nu_{e}}$ are the electron and neutrino rest 
masses, $p$ and $E$ are the electron momentum and kinetic energy, 
$W_{i}$ is the relative transition probability to the $i$th molecular final state 
of corresponding end-point energy $E_{0i}$. Fitting the  nuclear $\beta$-decay 
data with  Eq.~\ref{neuph}~ produces a significant negative 
value for the square of the electron neutrino mass \cite{Part94}.

We apply the metric deformation parameter $\epsilon$ obtained from the muon 
neutrino mass measurement to the nuclear $\beta$-decay and electron 
neutrino mass measurement. We do not lose on generality, but simplify our 
calculation, by assuming only one molecular final state.  In this case, 
the shape of the $\beta$-spectrum close to the end-point reduces to
\begin{equation}
W(E) \sim  p (E+m_{e})  (E_{0}-E) \sqrt{ (E_{0}-E)^{2} - m_{\nu_{e}}^{2}}.
\label{neuphr}
\end{equation}
If the decay happens in the deformed space ${\cal C}^{*}$, where, as in 
the case of the muon neutrino mass, we assume that the  ``true'' neutrino 
mass $m^{*2}_{\nu_{e}}=0$, then Eq.~\ref{neuphr}~ becomes
\begin{equation}
W(E^{*}) \sim p^{*} (E^{*}+m_{e}^{*}) (E_{0}-E^{*})^{2}.
\label{neuphd}
\end{equation}
All the kinematic quantities in the deformed space can be calculated from 
the quantities in the undeformed space using the transformation defined by 
the tensor  in  Eq.~\ref{dten}~ and the value of the parameter 
$\epsilon$ as determined from the ${\pi^{+}} \to {\mu^{+}}{\nu_{\mu}}$ decay.  
Thus $p^{*}=\epsilon p$, and, in a non-relativistic approximation, 
$E^{*} = {\epsilon}^{2} E$, and $m_{e}^{*} = m_{e}$.  We do not transform 
$E_{0}$ because it is a parameter in the distribution, and therefore a constant. 
Then,  Eq.~\ref{neuphd}~ become
\begin{equation}
W(E,\epsilon) \sim {\epsilon} p ( {\epsilon}^{2}E+m_{e}) 
(E_{0}- {\epsilon}^{2} E)^{2}.
\label{neuphd2}
\end{equation}
Assuming the weak-interaction volume has deformed metrics and 
the ``true'' neutrino mass is zero,  Eq.~\ref{neuphd2}~ represents 
the shape of the $\beta$-spectrum.  By assuming that the parameter 
$\epsilon$ is constant and that the transformation affects only momentum 
and not the total energy, as shown by  Eq.~\ref{dten}, 
we restrict ourselves to a very simple model. Even under these simplifying 
assumptions, there is sufficient new information contained in 
Eq.~\ref{neuphd2}~ that  we can study several experimental signatures 
resulting from this distribution:

\begin{itemize}  

\item We  verified that the shape of the distribution represented by 
Eq.~\ref{neuphd2}~ is consistent with existing measurements.
In Fig 1. we plot the deviation of this distribution  for $\epsilon = 0.999988$ 
from the distribution for  $\epsilon = 1$, corresponding to undeformed space, 
normalized to the undeformed distribution. It is clear that over the entire 
electron energy spectrum, except for the region very close to the end-point, 
the deviation is sufficiently small that it could not have been observed given 
the precision of existing experimental  measurements. 
This is even more true for $\epsilon = 0.9999998$. 

\item We plot the distribution in the region close to the end-point in Fig. 2, 
where we show that the values corresponding to $\epsilon = 0.999988$ lay 
above the undeformed distribution, consistent with the
distribution resulting in $m_{\nu _e}^{2} < 0$. 

\item If we assume that the electron energy distribution is described by 
Eq.~\ref{neuphd2}, but is fitted by the distribution with a shape described by
Eq.~\ref{neuph} , there would be a mismatch at some point in the spectrum. 
In Fig. 3 one can easily see this mismatch close the end-point for the case 
when the distribution is generated by Eq.~\ref{neuphd2}~ 
with parameter $\epsilon = 0.999988$, and then fit with the distribution 
with shape described by Eq.~\ref{neuph}.  This mismatch results in a bump 
close to the end-point, as shown in Fig. 4, experimentally corresponding to 
an overestimation of the counting rate. This effect is observed in many 
electron neutrino mass measurements 
\cite{Part94,Weinheimer93,Belesev95,Stoeffl95,Lobashev98,Zuber98,Schmitz99}. 

\item Finally, it has also been observed experimentally that $m_{\nu_{e}}^{2}$ is 
dragged further below the endpoint as a function of the lower limit of $E$ in 
the fit interval \cite{Weinheimer93,Lobashev98}. The $W(E)$ distribution 
generated using Eq.~\ref{neuphd2}~ with parameter $\epsilon = 0.999988$ 
and then fit  using the distribution with shape described Eq.~\ref{neuph}  
is plotted in Fig. 5, showing that the same effect is observed; 
namely that $m_{\nu_{e}}^{2}$ becomes more negative as the fit interval
is increased by  lowering the limit of $E$.

\end{itemize}

\begin{figure}
\begin{center}
\mbox{\epsfxsize=3.5in\epsffile{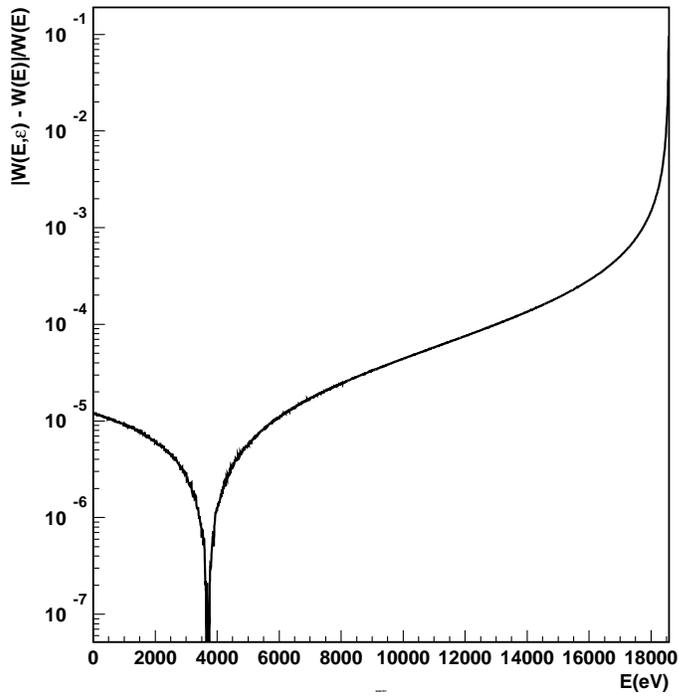}}
\end{center}
\caption{ Normalized deviation of the $\beta$-spectrum described by Eq.~\ref{neuphd2}~ for 
$\epsilon = 0.999988$ from an undeformed electron energy distribution. The deviation for 
$\epsilon = 0.9999998$ is much smaller.}
\end{figure}

\begin{figure}
\begin{center}
\mbox{\epsfxsize=3.5in\epsffile{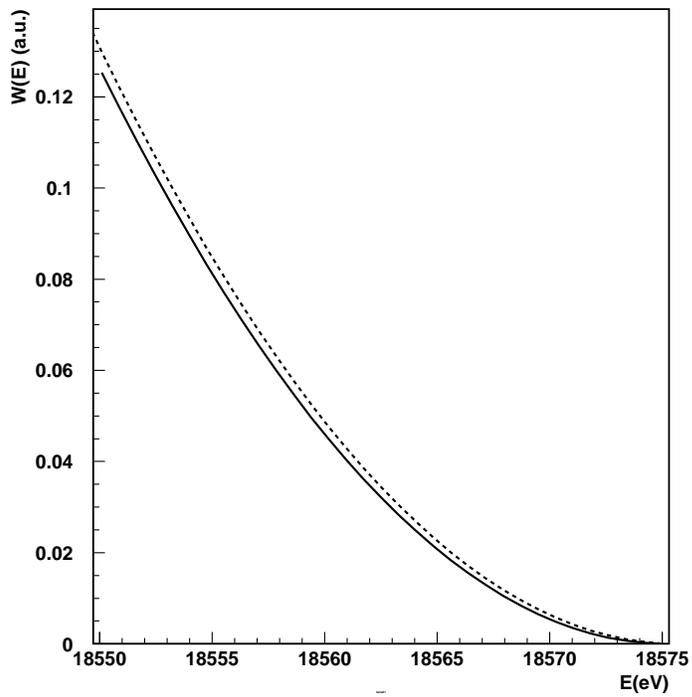}}
\end{center}
\caption{Beta spectrum near the end-point. The solid line is for an undeformed spectrum, 
while the dashed line 
represents the deformed spectrum for $\epsilon = 0.999988$.The deviation for 
$\epsilon = 0.9999998$ is less then the thickness of the line.}
\end{figure}

\begin{figure}
\begin{center}
\mbox{\epsfxsize=3.5in\epsffile{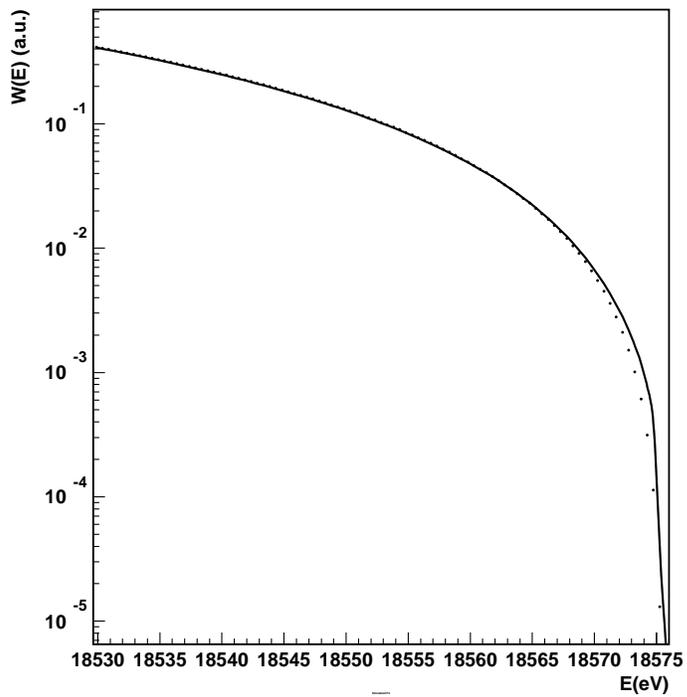}}
\end{center}
\caption{A mismatch close the end-point when the spectrum generated by Eq.~\ref{neuphd2}~ 
with parameter $\epsilon = 0.999988$ (dots) is fit with the distribution with shape described 
by Eq.~\ref{neuph}~ (solid line).}
\end{figure}

\begin{figure}
\begin{center}
\mbox{\epsfxsize=3.5in\epsffile{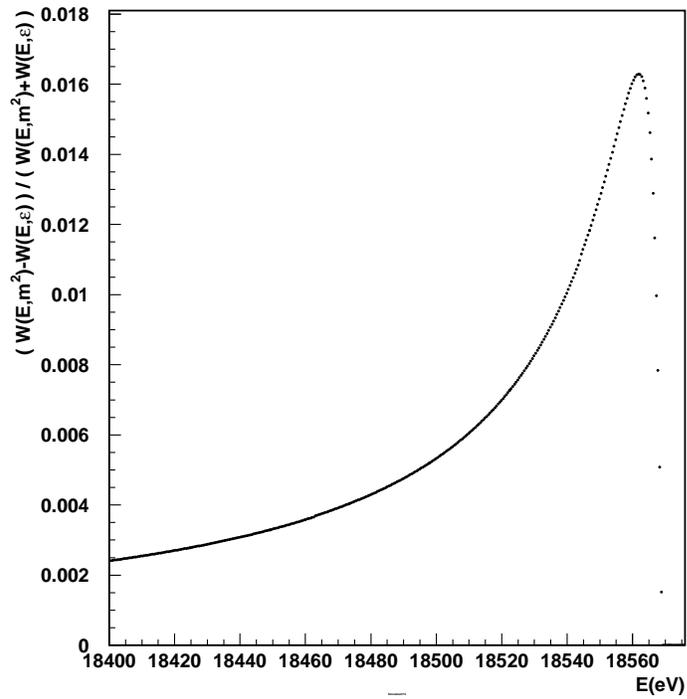}}
\end{center}
\caption{The bump produced close to the end-point by the fitting procedure described in the text,
experimentally corresponding  to an excess in counting rate close to the end-point. 
This is just a different presentation of the effect presented in Fig. 3.}
\end{figure}

\begin{figure}
\begin{center}
\mbox{\epsfxsize=3.5in\epsffile{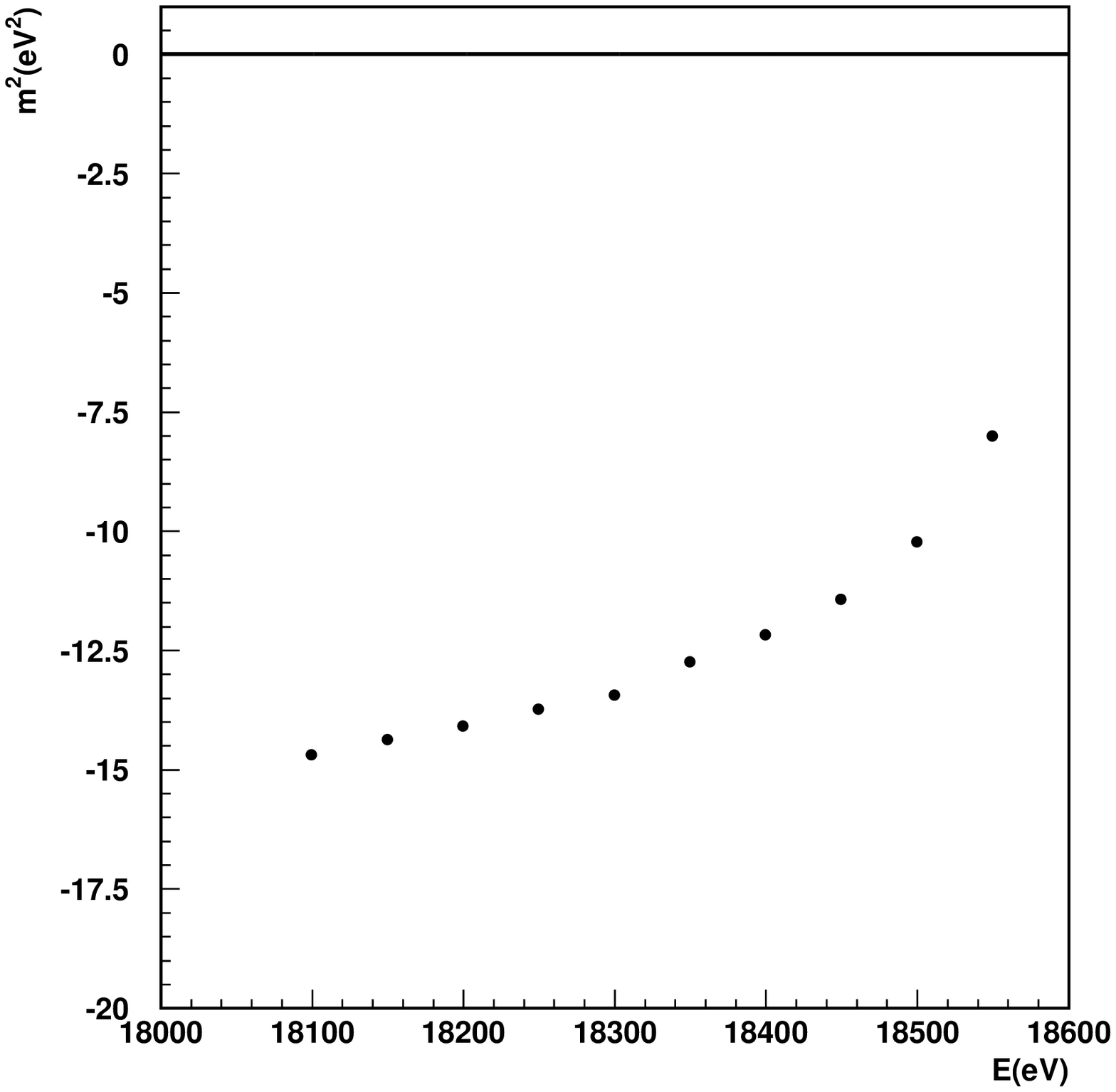}}
\end{center}
\caption{\label{f:overview1} Square of the neutrino rest mass $m_{\nu_{e}}^{2}$ obtained by 
fitting the $\beta$-spectrum generated with  Eq.~\ref{neuphd2}~ with $\epsilon = 0.999988$ as 
a function of the lower limit $E$ of the fit interval.}
\end{figure}

A more realistic deformation of space, in which there would be more then one 
free parameter, and whose parameters could be energy and momentum 
dependent (generally as in Eq. 15), would result in a different relation in 
the mass-energy equation, where the calculation would be more complex 
and harder to relate to existing experimental results. Finding the exact 
deformation function is far beyond the scope of this paper, and our model is 
made as simple as possible. Despite its simplicity, this model still results in
several significant consequences which already have been or could be 
experimentally tested. 

\section{Conclusion}

In this paper we have suggested, analogous to  the general
theory of relativity and the 
theory of deformation for continuous media, a simple mechanism in which 
the negative values  for the square of the neutrino mass reported in most
of the neutrino experiments \cite{Part94} could be the result of a change
in metrics in the small weak interaction volume
in the energy-momentum representation. We constructed a simple model in
which the changes in energy-momentum metrics do result in 
$m^{2}_{\nu} \leq 0$, while at the same time no components
of the model violated allowed physical limits. The goal of this paper was not to
construct the complete theory of metric deformation, but rather
to demonstrate that the negative value of the square of the neutrino mass should
not immediately be discarded as unphysical, and could indicate 
new physical phenomena.
 
I would like to thank Steven P. Wells for long and useful discussions.

\newpage

\end{document}